\documentclass[pre,twocolumn,showpacs,floatfix,superscriptaddress]{revtex4}

\usepackage{listings}
\usepackage{graphicx}
\usepackage{epstopdf}
\usepackage{multirow}
\usepackage{color}
\usepackage{amssymb,amsfonts,amsmath}
\usepackage[english]{babel}
\usepackage{float}

\begin{document}


\title{How citation boosts promote scientific paradigm shifts and Nobel Prizes}

\author{Amin Mazloumian}\affiliation{ETH Z\"urich, CLU E1, Clausiusstrasse 50,
  8092 Z\"urich, Switzerland}

\author{Young-Ho Eom}\affiliation{Complex Networks \& Systems Lagrange Laboratory, ISI Foundation, Turin, Italy}

\author{Dirk Helbing}\affiliation{ETH Z\"urich, CLU E1, Clausiusstrasse 50,
  8092 Z\"urich, Switzerland}

\author{Sergi Lozano}\affiliation{ETH Z\"urich, CLU E1, Clausiusstrasse 50,
  8092 Z\"urich, Switzerland}

\author{Santo Fortunato}\affiliation{Complex Networks \& Systems Lagrange Laboratory, ISI Foundation, Turin, Italy}

\begin{abstract}
Nobel Prizes are commonly seen to be among the most prestigious achievements of our times.
Based on mining several million citations, we quantitatively analyze the processes driving paradigm shifts in science. We find that groundbreaking discoveries of Nobel Prize Laureates and other famous scientists are not only acknowledged by many citations of their landmark papers. Surprisingly, they also boost the citation rates of their previous publications. Given that innovations must outcompete the rich-gets-richer effect for scientific citations, it turns out that they can make their way only through citation cascades. A quantitative analysis reveals how and why they happen. Science appears to behave like a self-organized critical system, in which citation cascades of all sizes occur, from continuous scientific progress all the way up to scientific revolutions, which change the way we see our world. Measuring the ``boosting effect'' of landmark papers, our analysis reveals how new ideas and new players can make their way and finally triumph in a world dominated by established paradigms. The underlying "boost factor" is also useful to discover scientific breakthroughs and talents much earlier than through classical citation analysis, which by now has become a widespread method to measure scientific excellence, influencing scientific careers and the distribution of research funds. Our findings reveal patterns of collective social behavior, which are also interesting from an attention economics perspective. Understanding the origin of scientific authority may therefore ultimately help to explain, how social influence comes about and why the value of goods depends so strongly on the attention they attract. 
\end{abstract}


\maketitle

\section{Introduction}

Ground-breaking papers are extreme events~\cite{albeverio06} in science.
They can transform the way in which researchers do science in terms of the subjects they choose, the methods
they use, and the way they present their results. The related 
spreading of ideas has been described as an epidemic percolation process in a social 
network~\cite{bettencourt06}. However, the impact of most innovations
is limited. There are only a few ideas, 
which gain attention all over the world and across disciplinary boundaries~\cite{davenport01}. Typical
examples are elementary particle physics, the theory of evolution, superconductivity, 
neural networks, chaos theory, systems biology, nanoscience, or network theory. 
\par
It is still a puzzle, however, how a new idea and its proponent can be successful, 
given that they must beat the rich-gets-richer dynamics of already established ideas and scientists. 
According to the Matthew effect~\cite{merton68, merton88,scharnhorst97,petersen11}, 
famous scientists receive an amount of credit that may sometimes appear disproportionate
to their actual contributions, 
to the detriment of younger or less known scholars. This implies a great authority of a small number of scientists, which is reflected by the big attention received by their work and
ideas, and of the scholars working with them~\cite{malmgren10}.
\par
Therefore, how can a previously unknown scientist establish at all a high scientific reputation and authority,  if those who get a lot of citations receive even more over time? Here we shed light on this puzzle. The following results for $124$ Nobel Prize Laureates in chemistry, economics, medicine and physics suggest that innovators can gain reputation and innovations can successfully spread, mainly {\it because} a scientist's body of work overall enjoys a 
greater impact after the publication of a landmark paper. Not only do
colleagues notice the ground-breaking paper, but
the latter also attracts the attention to older publications of the same 
author (see Fig. 1). 
\begin{figure} [h]
\begin{center}
\includegraphics[width=\columnwidth]{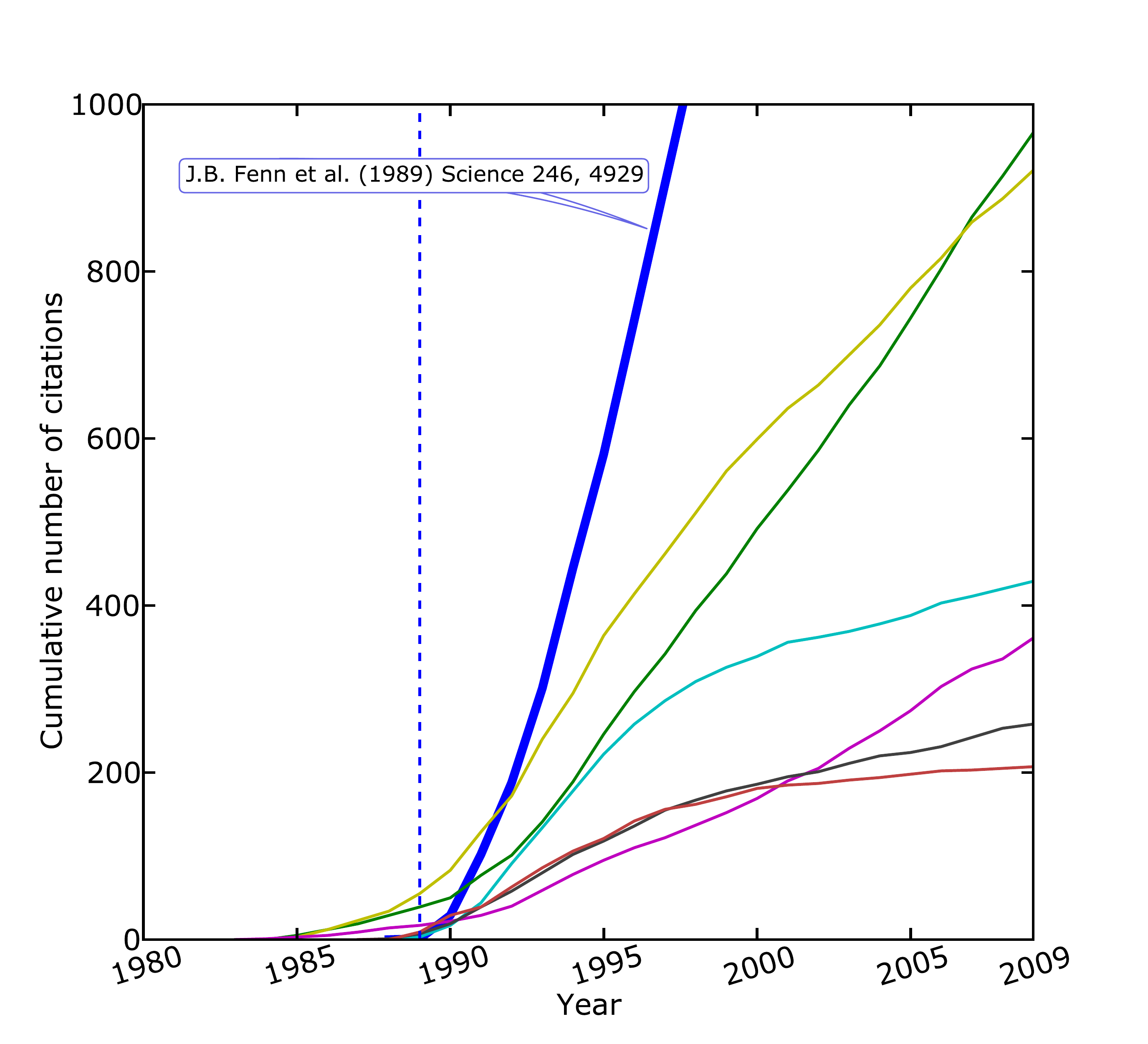}
\caption{Illustration of the boosting effect.
Typical citation trajectories of papers, here for Nobel Prize
Laureate John Bennett Fenn, who received the award in chemistry in 2002
for the development of the electrospray
ionization technique used to analyze biological
macromolecules. The original article, entitled 
{\it Electrospray ionization for mass spectrometry of large
biomolecules}, coauthored by M. Mann, C. K. Meng, S. F. Wong and 
C. M. Whitehouse, was published in $\it Science$ in 1989 and is the most
cited work of Fenn, with currently over $3,000$ citations. The
diagram reports the growth in time of the total number of citations
received by this landmark paper (blue solid line) and by six older
papers. 
The diagram indicates that the number of citations of the landmark paper has literally exploded in the
first years after its appearance. However, after its publication in 1989, a number of
other papers also enjoyed a much higher citation rate. Thus, a
sizeable part of previous scientific work 
has reached a big impact after the publication of the landmark
paper.  We found that the occurrence of this boosting effect is characteristic for successful scientific careers.}
\end{center}
\end{figure}
Consequently, {\it future} papers have an impact on {\it past} papers,
as their relevance is newly weighted. 
\par
We focus here on citations as indicator of scientific 
impact~\cite{garfield55,garfield79,egghe90,amsterdamska89,petersen10}, studying data from the ISI Web of
Science, but the use of click streams~\cite{bollen05} would be conceivable as well. It is
well-known that the relative number of citations correlates with research 
quality~\cite{trajtenberg90,aksnes06,moed05}. Citations are now
regularly used in university rankings~\cite{vanraan05},
in academic recruitments and for the distribution of funds among scholars and scientific
institutions~\cite{boyack03}. 

\section{Results}

We evaluated data for $124$ Nobel Prize Laureates that were 
awarded in the last two decades ($1990$-$2009$), which include an impressive
number of about $2$ million citations. For all of them and
other internationally established experts as well, 
we find peaks in the changes of their citation rates (Figs. 2 and 3). 
\begin{figure} [h]
\begin{center}
\includegraphics[width=\columnwidth]{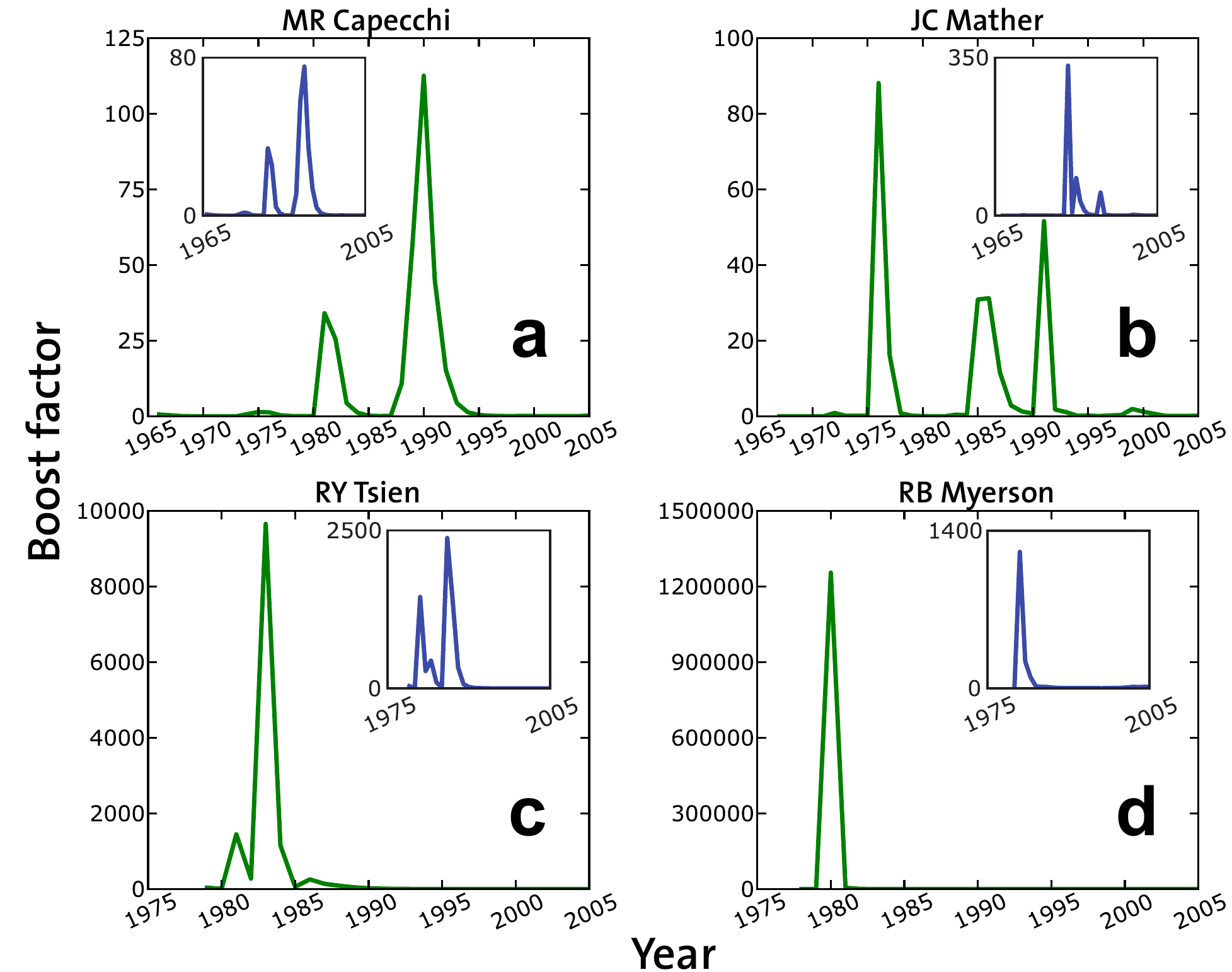}
\caption{Typical time evolutions of the boost
      factor. Temporal dependence of $R'_w(t)$ for Nobel
Laureates [here for (a) Mario R. Capecchi (Medicine, 2007), 
  (b) John C. Mather (Physics, 2006), 
(c) Roger Y. Tsien (Chemistry, 2008) and (d) Roger B. Myerson (Economics, 2007)]. 
Sharp peaks indicate citation boosts in favor of older papers, 
triggered by the publication and recognition of
a landmark paper. Insets: The peaks even persist (though somewhat smaller), if in the determination of
the citation counts $c_{p,t}$, the landmark paper is skipped (which is defined as the paper that produces 
the largest reduction in the peak size, when excluded from the computation of the boost factor).
We conclude that the observed citation boosts are mostly due to a collective effect
involving several publications rather than due to the high citation rate
of the landmark paper itself.}
\end{center}
\end{figure}
\begin{figure} [h]
\begin{center}
\includegraphics[width=\columnwidth]{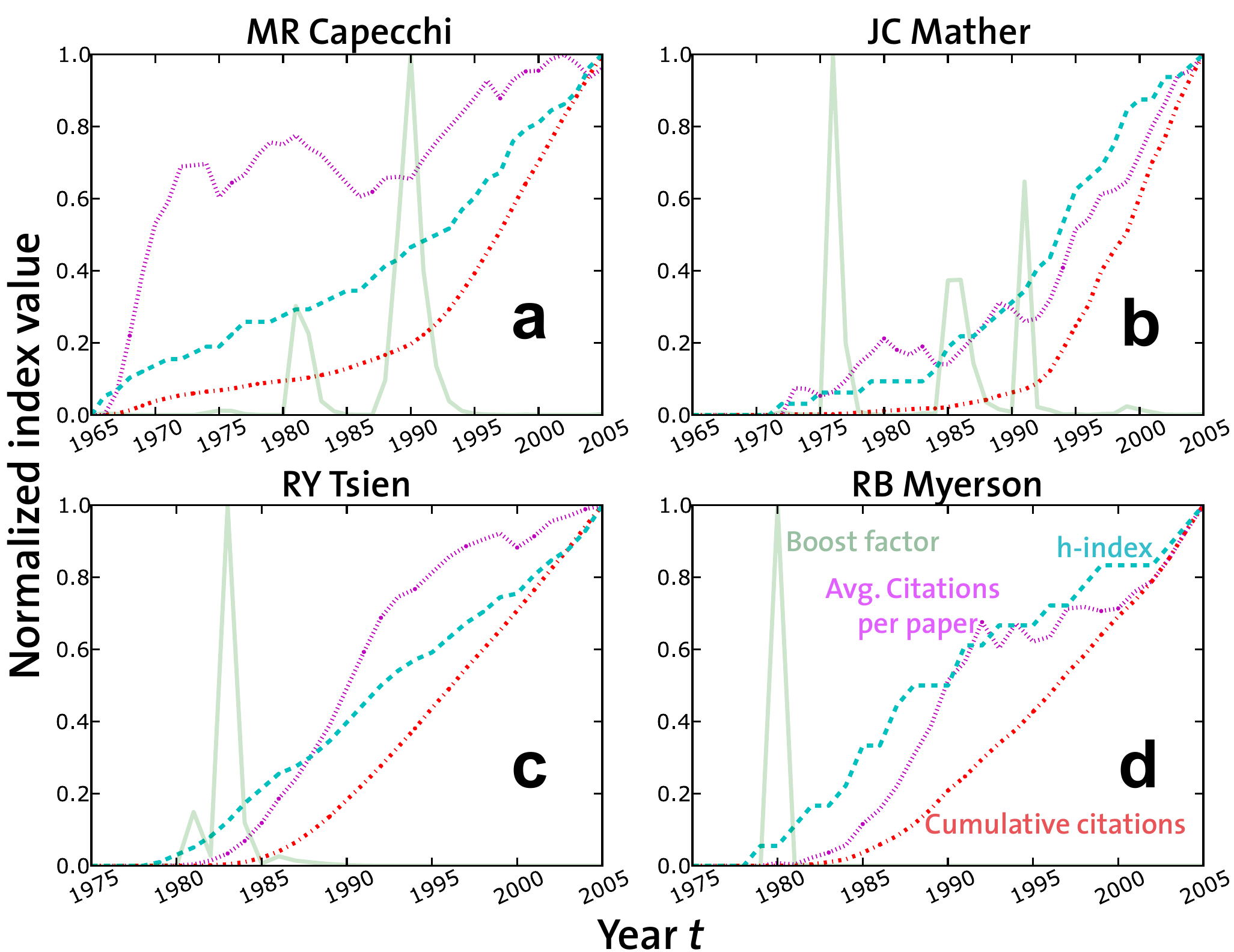}
\caption{Dynamics of the boost factor $R'_w(t)$ versus traditional
citation variables. Each panel displays the time histories of four
variables: the boost factor $R'_w(t)$, the average number of citations per
paper $\langle c(t)\rangle$, the cumulative number of citations $C(t)$, and the 
$H$-index earned until year $t$~\cite{hirsch05}. The panels refer to the same Nobel Laureates as displayed in
Fig. 2. The classical indices have relatively smooth profiles, i.e. they are
not very sensitive to extreme events in the life of a scientist like
the publication of landmark papers. An advantage of the boost factor is that its peaks allow one to
identify scientific breakthroughs earlier.}
\end{center}
\end{figure}

Moreover,
it is always possible to attribute to these peaks landmark
papers (Fig. 4), which have reached hundreds of citations over the
period of a decade. 
\begin{figure} [h]
\begin{center}
\includegraphics[width=\columnwidth]{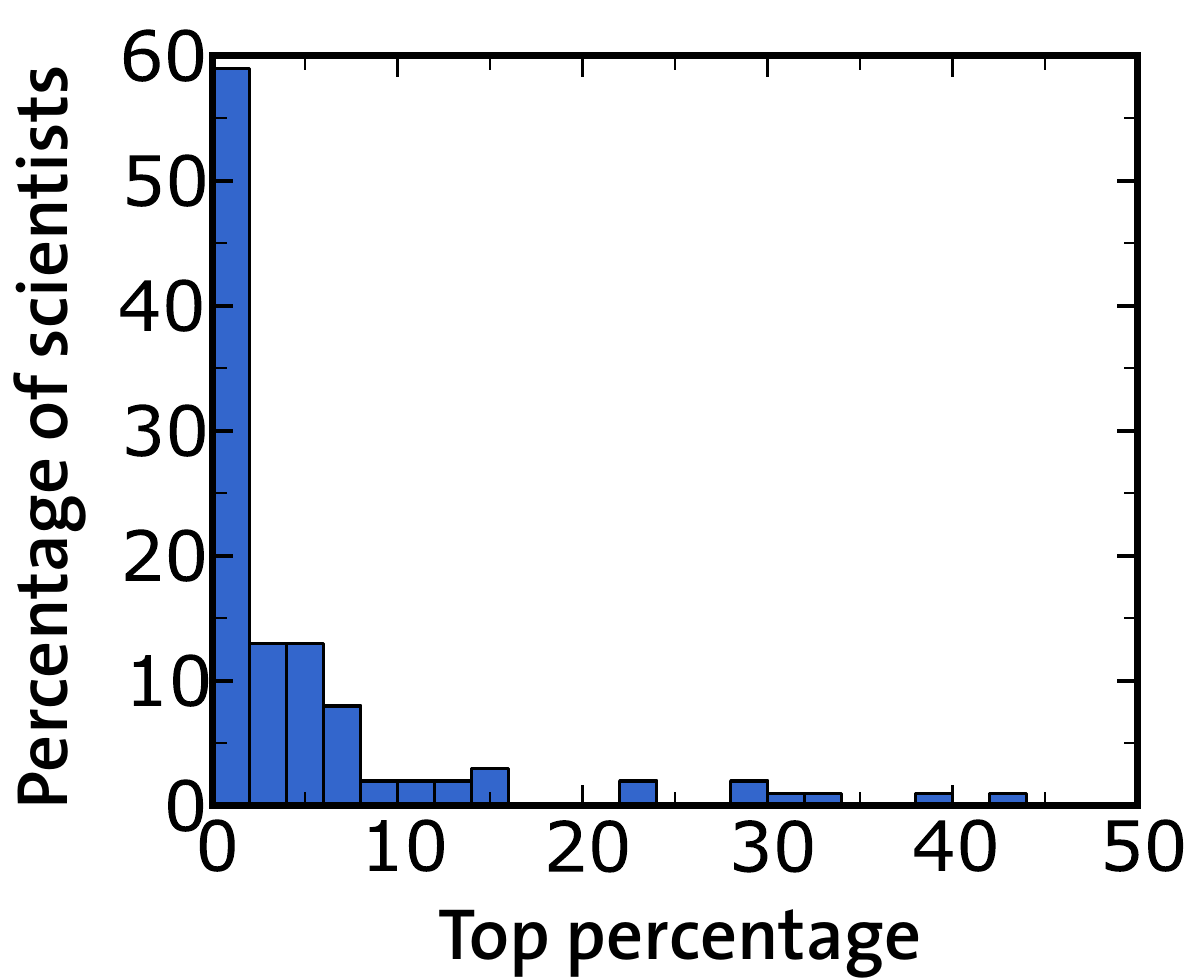}
\caption{Correlation between papers and the local maxima (``peaks'') of $R'_w(t)$.
We first determined the ranks of all papers of an author based on
the total number of citations received until the year $2009$ inclusively. We then
determined the rank of that particular publication, which had the greatest
contribution to the peak. This was done by measuring the reduction
in the height of the peak, when the paper was excluded from the
calculation of the boost factor (as in the insets of Fig. 2). 
The distribution of the ranks of ``landmark papers'' is dominated by low values,
implying that they are indeed among the top publications of their
authors.}
\end{center}
\end{figure}
Such landmark papers are rare even in
the lives of the most excellent scientists, 
but some authors have several such peaks. 
\par
Technically, we detect a groundbreaking article $a$ published at time
$t=t_a$ by comparing the citation rates before and after $t_a$ for the
earlier papers. 
The analysis proceeds as follows:  Given a year $t$ and a time window $w$, we take all papers
of the studied author that were published since the beginning of his/her career
until year $t$. The citation rate $R_{<t,w}$ measures the average number
of citations received per paper per year in the period from $t-w+1$ to
$t$. Similarly, 
the citation rate $R_{>t,w}$ measures the average number of citations
received by the same publications per paper per year between $t+1$ and $t+w$ (or
$2009$, if $t+w$ exceeds $2009$). The ratio
$R_w(t)=R_{>t,w}/R_{<t,w}$, which we call the 
``boost factor'', is a variable that detects critical events in the life of a scientist: sudden increases in
the citation rates (as illustrated by Fig. 1) show up as peaks in the time-dependent plot of $R_w(t)$. 
\par
In our analysis we used the generalized boost factor $R'_w(t)$, 
which reduces the influence of random variations in the citation rates
(see Materials and Methods).

Figure 2 shows typical plots of the boost factors $R'_w(t)$ of four Nobel Prize Laureates. 
Interestingly, peaks are even found, when those papers, which mostly contribute to them, 
are {\it excluded} from the analysis (see insets of Fig. 2). That is,
the observed increases in the citation rates are not just due to the landmark papers themselves, but
rather to a collective effect, namely an 
increase in the citation rates of \textit{previously} published
papers. This results from the greater visibility that the body of work of the corresponding
scientist receives after the publication of a landmark paper and establishes an increased 
scientific impact (``authority''). From the perspective of attention economics~\cite{wu07}, it may be interpreted 
as a herding effect resulting from the way in which relevant information is collectively discovered
in an information-rich environment. Interestingly, we have found
that older papers
receiving a boost are not always works related to the topic of the
landmark paper.
\par
Traditional citation analysis does not reveal such crucial
events in the life of a scientist very well. Figure~3
shows the time history of three classical citation indices:
the average number of citations per
paper $\langle c(t)\rangle$, the cumulative number $C(t)$ of citations,
and the Hirsch index~\cite{hirsch05} ($h$-index) $H(t)$ in year
$t$. For comparison, the evolution of the boost factor $R'_w(t)$ is depicted as well.
All indices were divided by their maximum value, in order to normalize
them and to use the same scale for all.
The profiles of the classical indices are rather smooth in most
cases, and it is often very hard to see any significant effects of  
landmark papers. However, this is not surprising, as the boost
  factor is designed to capture abrupt variations in the citation
  rates, while both $C(t)$ and $H(t)$ reflect the overall
  production of a scientist and are therefore less sensitive to extreme events.
\par
To gain a better understanding of our findings, Figs. 4 and 5 present a statistical analysis
of the boosts observed for Nobel Prize Laureates. 
\begin{figure} [h]
\begin{center}
\includegraphics[width=\columnwidth]{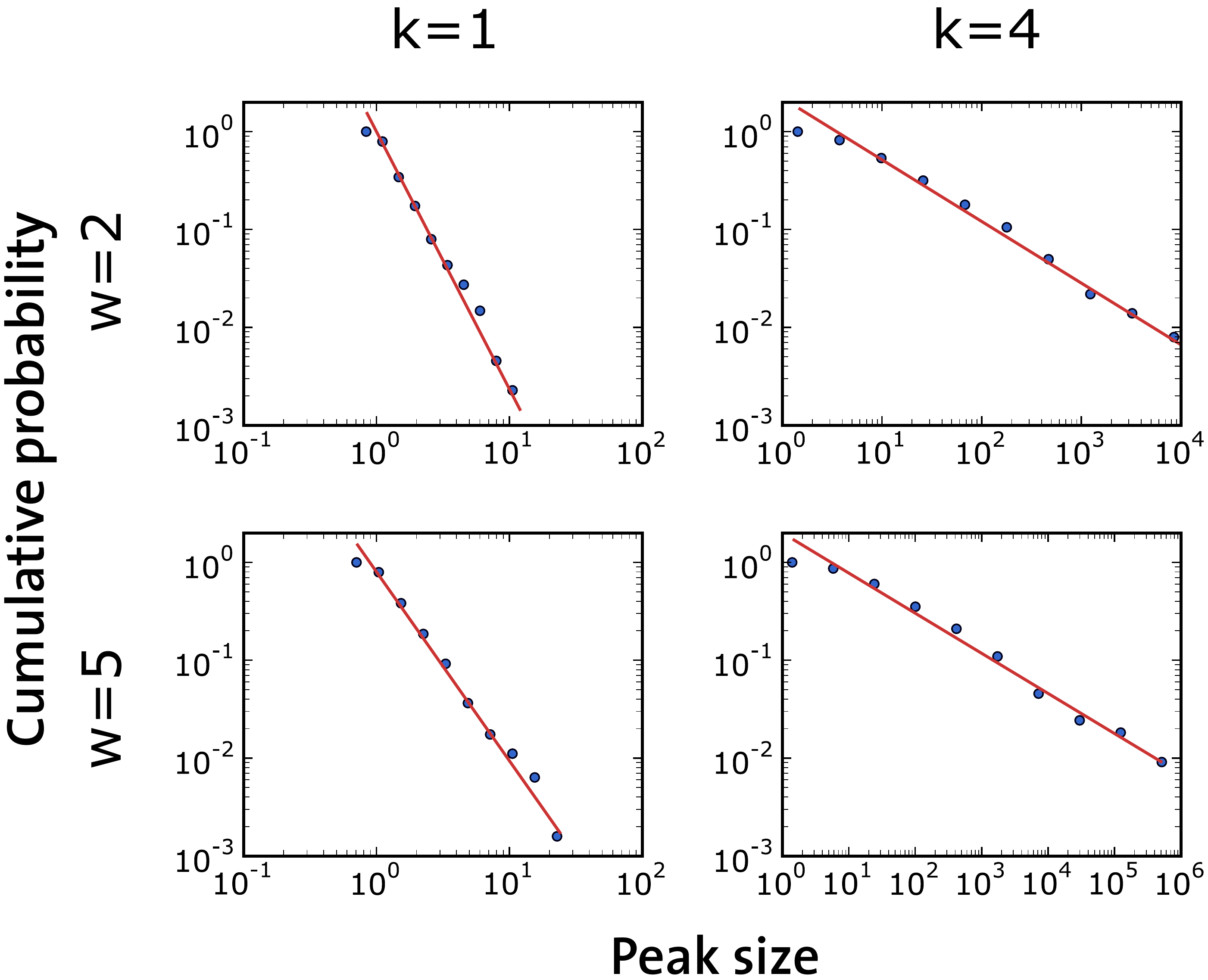}
\caption{Cumulative probability distribution of peak heights in the boost factor
curves of Nobel Prize Laureates. The four panels correspond to
different choices of the parameters $k$ and $w$. The power law fits
(lines) are performed with the maximum likelihood
method~\cite{clauset07}. The
exponents for the direct distribution (of which the
cumulative distribution is the integral) are: $3.63\pm 0.16$ (top left),
$2.93\pm 0.16$ (bottom left), $1.63\pm 0.05$ (top right), $1.41\pm 0.05$ (bottom
right). The best fits have the following lower cutoffs and values of
the Kolmogorov-Smirnov (KS) statistics: $1.06$, $0.0289$ (top left),
$1.15$, $0.0264$ (bottom left), $13.1$, $0.038$ (top right),
$24.7$, $0.0462$ (bottom right). The KS values support the power law
ansatz for the shape of the curves. Still, we point out that on the
left plots the data span just one decade in the variable, so one has
to be careful about the existence of power laws here.}
\end{center}
\end{figure}
Figure 4 demonstrates that
pronounced peaks are indeed related to highly cited papers. Furthermore,
Fig. 5 analyzes the size distribution of peaks. The distribution
looks like a power law for all choices of the parameters 
$w$ and $k$ (at least within the relevant range of small values).
This suggests that the bursts are produced by citation cascades as they
would occur in a self-organized critical system~\cite{bak87}. In fact, power laws were found
to result from human interactions also in other contexts~\cite{barabasi05,oliveira05,malmgren09}. 
\par
The mechanism underlying citation cascades is the discovery of new ideas, which colleagues refer to 
in the references of their papers. Moreover, according to the rich-gets-richer effect, 
successful papers are more often cited, also to raise their own success. Innovations 
may even cause scientists to change their research direction or approach. Apparently, 
such feedback effects can create citation cascades, which are ultimately triggered by landmark papers.

Finally, it
  is important to check whether the boost factor is able to distinguish
  exceptional scientists from average ones. Since any criteria used to
  define ``normal scientists'' may be questioned, we have assembled a
  set of scientists taken at random. Scientists were chosen among those who published at least one
paper in the year 2000. We selected $400$ names 
  for each of four fields: Medicine, Physics, Chemistry and
  Economy. After discarding those with no citations, we ended up with $1361$
scientists. In Fig. 6 we draw on a bidimensional plane each scientist of our random
sample (empty circles), together with the Nobel Prize Laureates 
considered (full circles). The two dimensions are the value of
the boost factor and the average number of citations of a
scientist. A cluster analysis separates the populations in the
proportions of $79\%$ to $21\%$. The separation is significant but
there is an overlap of the two datasets, mainly because of
two reasons. First, by picking a large number of scientists at
random, as we did, there is a finite
probability to choose also outstanding scholars. We have
verified that this is the case. Therefore, some of the empty circles
deserve to sit on the top-right part of the diagram, like many Nobel
Prize Laureates. The second reason is that we are considering
scholars from different disciplines, which generally have different citation
frequencies. This affects particularly the
average number of citations of a scientist, but also the value of the
boost factor. In this way, the position in the diagram is
affected by the specific research topic, and the distribution of the
points in the diagram of Fig. 6 is a superposition of field-specific distributions. Nevertheless, the two datasets,
though overlapping, are clearly distinct. Adding further dimensions could
considerably improve the result. In this respect, the boost factor can
be used together with other measures to better specify the performance
of scientists.
\begin{figure} [h]
\begin{center}
\includegraphics[width=\columnwidth]{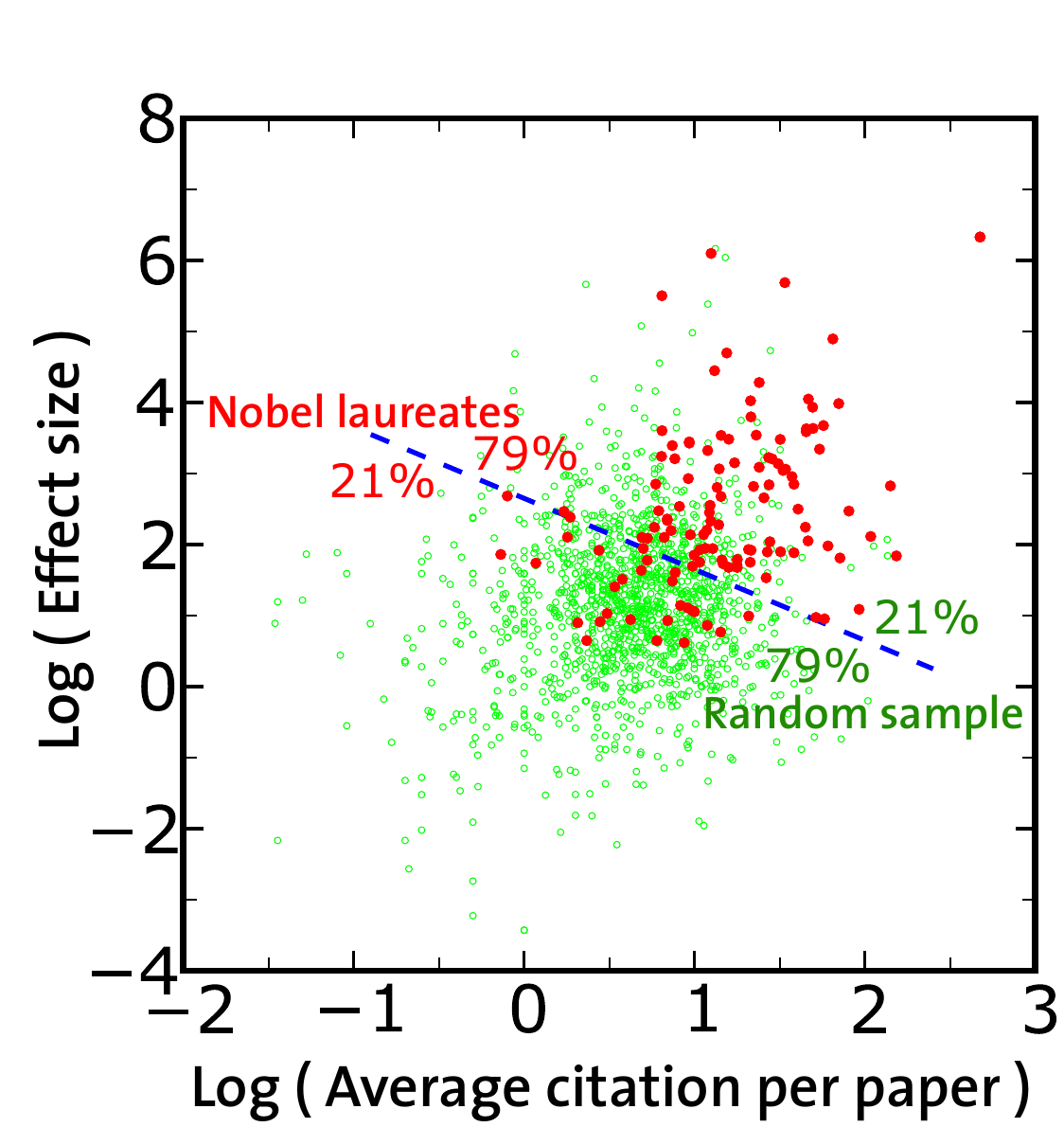}
\caption{Two-dimensional representation of our collection of
  Nobel Prize Laureates and a set of 1361 scientists, which were randomly
  selected. On the x-axis we report the average number of citations of
a scientist, on the y-axis his/her boost factor. It can be seen that,
on average, Nobel Prize winners clearly perform better. However a Nobel
Prize is not solely determined by the average number of citations and
the boost factor, but also by further factors. These may be the degree
of innovation or quality, which are hard to quantify.}
\end{center}
\end{figure}

\section{Discussion}

In summary, groundbreaking scientific papers have a boosting effect on
previous publications of their authors, bringing them to the attention of the
scientific community and establishing their ``authority''. We have
provided the first quantitative characterization of this phenomenon 
by introducing a new variable, the ``boost factor'',
which is sensitive to sudden changes in the citation rates. 
The fact that landmark papers trigger the collective discovery of older papers amplifies their impact and tends to generate pronounced spikes long before the paper
receives full recognition. The boosting factor can therefore serve to discover
new breakthroughs and talents more quickly than classical citation indices. It may also 
help to assemble good research teams, which have a pivotal role in modern
science~\cite{guimera05,wuchty07,jones08}. 
\par
The power law behavior observed in
the distribution of peak sizes suggests that science progresses
through phase transitions~\cite{stanley87} with citation avalanches on all scales---from small cascades 
reflecting quasi-continuous scientific progress all the way up to scientific revolutions, which 
fundamentally change our perception of the world. While this provides new
evidence for sudden paradigm shifts~\cite{kuhn62}, our results also give a
better idea of why and how they happen.
\par
It is noteworthy that similar feedback effects may determine
the social influence of politicians, or prices of stocks and products (and, thereby, the value of companies).
In fact, despite the long history of research on these subjects, such phenomena
are still not fully understood. There is evidence, however, that the power of a person
or the value of a company increase with the level 
of attention they enjoy. Consequently, our study of
scientific impact is likely to shed new light on these scientific puzzles as well.

\section{Materials and Methods}
The basic goal is to improve the signal-to-noise ratio in the citation rates, in order to detect 
sudden changes in them. An effective method to reduce the influence of papers with largely fluctuating citation rates is to weight highly cited papers more.
This can be achieved by raising the number of cites to the power $k$,
where $k>1$. Therefore, our formula to compute $R'_w(t)$ looks as follows:
\begin{equation}
R'_w(t)=\frac{\sum_p\sum_{t^\prime=t+1}^{t+w} (c_{p,t^\prime})^k}{\sum_p\sum_{t^\prime=t-w+1}^{t} (c_{p,t^\prime})^k}.
\label{eq1}
\end{equation}
Here, $c_{p,t^\prime}$ is the number of cites received by paper $p$ in
year $t^\prime$. The sum over $p$ includes all papers published before
the year $t$; $w$ is the time window selected to compute the boosting effect.
For $k=1$ we recover the original definition of $R_w(t)$ (see main
text). For the analysis presented in the paper we have used $k=4$ and
$w=5$, 
but our conclusions are not very sensitive to the choice of smaller values of $k$ and $w$.

\section{Acknowledgments}

We acknowledge the use of ISI Web of Science data of Thomson Reuters for our citation analysis. 
A.M., S.L. and D.H. were partially supported by the Future and Emerging Technologies programme FP7-COSI-ICT of the European Commission through the project QLectives (grant no.: 231200). Y.-H. E. and S. F. gratefully acknowledge ICTeCollective, grant 238597 of the European Commission.

\end{document}